\documentclass[a4paper,11pt]{article}
\usepackage{pos}
\usepackage{epsfig}
\usepackage{amsmath}
\usepackage{color}

\usepackage{graphicx}
\title{Electroweak Physics: Summary}
\ShortTitle{Electroweak Physics}

\author*[a]{Rupa Chatterjee}
\affiliation[a]{Variable Energy Cyclotron Centre,  1/AF, Bidhan Nagar, Kolkata-700064, India}

\emailAdd{rupa@vecc.gov.in}
\abstract{Electroweak probes are  potential tool to study the properties of the hot and dense strongly interacting matter produced in relativistic nuclear collisions due to their unique nature. A selection of the new experimental analysis and results from theory calculations on electromagnetic and weak probes presented at the Hard Probes 2020 are discussed in this contribution.}

\FullConference{%
  HardProbes2020\\
  1-6 June 2020\\
  Austin, Texas}

\begin{document}
\maketitle

%\section{Introduction} 
The Quark-Gluon Plasma (QGP) phase produced in relativistic nucleus-nucleus collisions is expected to have a transient existence and the information about the space-time evolution and dynamics this hot and dense matter are obtained from the various final state observables. Photons (real as well as  virtual) are emitted from the entire lifetime of the  matter evolution and the weak gauge bosons  are  produced from the initial hard scatterings. These electroweak probes do not suffer strong interaction with the medium, carry undistorted information from their production point to the detector and thus are regarded as efficient probes to study the initial state and the properties of the  strongly interacting  matter~\cite{mclaurren, gojkohp, zvi}.
%Electroweak probes, i.e., photons (both real and virtual) and weak gauge bosons are the most 

\section{Direct Photons}
Electromagnetic radiations are known as the thermometer of the medium from early days of heavy ion collisions and the direct photons were initially studied  to get the temperature of the system formed in these collisions~\cite{dks_qm}.

The experimentally measured inclusive photon spectrum contains a huge background that originates (mostly) from the 2-$\gamma$ decay of $\pi^0$ and $\eta$ mesons. The direct photon spectrum is obtained by subtracting this decay background from the inclusive photon spectrum. Photons produced from the various stages of the matter evolution contribute to the direct photon spectrum. %and it contains contributions from the various stages of the evolving fireball produced in heavy ion collisions. % of collision fireball%Photons produced from various stages of the evolving fireball contributes to the direct photon spectrum. % contains contribution from various stages of the evolution of collision fireball. %There are various sources of the direct photon production from heavy ion collisions. 
The prompt photons are produced from the initial hard scatterings and the pre-equilibrium photons are emitted before the medium gets thermalized. These photons dominate the high $p_T$ region of the direct photon spectrum. The thermal photons are radiated from the QGP as well as from the hot hadronic matter  and populate the  $p_T < 4$ GeV region of the transverse momentum spectrum. The jet-conversion process also contributes significantly to the direct photon spectrum. %The direct photons are also produced by jet conversion process. 

%The next to leading order perturbative QCD calculations have explained the the 

%The main difficulty in the experimental measurement of real photons is the huge background which mainly originates from the decay of  . It is to be noted that $\omega$ and $\eta^{\prime}$ also contributes the the decay background of the inclusive photon spectra.

There has been significant advancement in the decay background subtraction methods in last couple of decades. The WA98 Collaboration first used the invariant mass analysis method to subtract the decay background from the inclusive photon spectrum~\cite{wa98}.  In recent times the PHENIX Collaboration at RHIC and the ALICE Collaboration at the LHC  have used several sophisticated methods  for the subtraction of the decay background.  

The experimental data at RHIC and at the LHC  have shown an  excess of direct photon yield from heavy ion collisions compared to the (scaled) photon yield from p+p collisions in the region $p_T < 4$ GeV. This excess yield is considered as thermal radiation from the hot and dense medium~\cite{ph_g, al_g}. The direct photon anisotropic flow  has  been  measured at RHIC and LHC energies at different centrality bins~\cite{v2_ex, v2_ex1}. It is well known that the thermal radiation completely dominates the photon anisotropic flow parameter~\cite{phot_v2} and the contributions from other (non-thermal) sources to photon $v_n$ are negligible. However, the theoretical model calculations which explain the charged particle spectra and anisotropic flow successfully, have been found to underpredict the experimental data of photon anisotropic flow by a significant margin~\cite{puzzle}. This is known as direct photon puzzle. %TheHowever, that the anisotropic flow reported by PHENIX from Au+Au collisions at RHIC does not change significantly with this new improvement. 

Direct photon data from a number of collision systems have been reported by the PHENIX Collaboration at RHIC in recent times.  Apart from the Au+Au and Cu+Cu collisions at 200A GeV,  photons  from 62.4 and 39A GeV Au+Au collisions have been measured  by PHENIX~\cite{veronika, phot_scaling}. Photon data from p+p, p+Au, and d+Au collisions at 200 A GeV are also available now. The 0--5\%  p+Au collisions have indicated a thermal production from the medium among the small systems at RHIC.
The PHENIX Collaboration has  also reported a new analysis method for the direct photon measurement where the double ratio $R_\gamma$ is calculated using external conversion (that leads to better cancellation of systematics). The results from the new analysis are found to be consistent with the $R_\gamma$ obtained from the internal conversion, virtual, as well as calorimeter methods~\cite{veronika}.% are % However, that the anisotropic flow reported by PHENIX from Au+Au collisions at RHIC does not change significantly with this new improvement. 

One important observation in recent time is the universal scaling behaviour of the low $p_T$ photon yield which scales with the charged particle multiplicity as $(dN_{\rm ch}/d\eta)^\alpha$~\cite{phot_scaling}. The scale factor $\alpha$ (is about 1.25) indicates that the photon multiplicity grows faster than the charged particle multiplicity. The scaling behaviour is found to be independent of the energy, centrality, and system size (see Fig. 1). The STAR data also show a similar scaling however, the magnitude is smaller than PHENIX. 

It is to be noted that a theory calculation by Cleymans ${\it et \ al.}$~\cite{dks} predicted similar scaling behaviour between photons and charged particle multiplicity long before where they estimated the scaling coefficient to be about 1.2, close to the $\alpha$ value by PHENIX. % The value of $\alpha$ by PHENIX is  The scaling coefficient by PHENIX is found to close to the value predicted by Claymans {\it et al.}
% scaling relation from the the photon production in heavy ion collision is the 
%New data set and theory results may help understanding this puzzle.

New prompt photon data at LHC  have been reported by the ALICE Collaboration. Isolated prompt photons from 7 TeV p+p collisions are available in the range 10 $< \, p_T \, < 60$ GeV where an isolation cut is used to reduce the fragmentation photons. The data is found to be in good agreement  with the JETPHOX NLO pQCD calculations~\cite{Dhrub}. %The photon data from 530 GeV p+p collisions  have also been explained successfully by the NLO pQCD calculation.
\begin{figure}
\centerline{\includegraphics*[width=11.0 cm, clip=true]{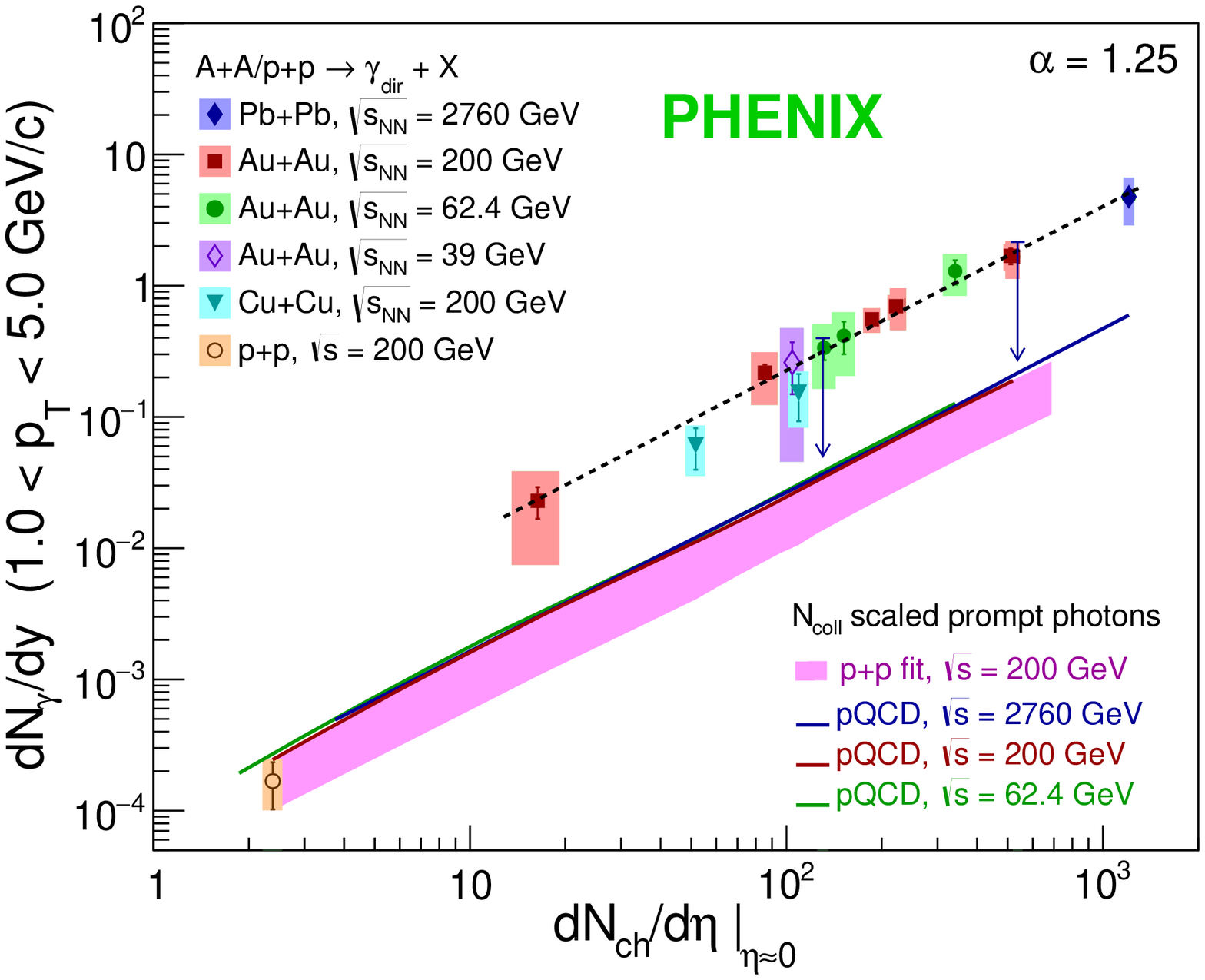}}
%\centerline{\includegraphics*[width=10.0 cm, clip=true]{2.ps}}
\caption{(Color online) Integrated  direct  photon  yield  ($1.0 < p_T < 5.0$  GeV) scales with the charged particle multiplicity $dN_{\rm ch}/d\eta$. The figure is from Ref.~\cite{phot_scaling}.}
\label{fig3}
\end{figure}

There has been significant advancement in the theory calculation of direct photon spectra and anisotropic flow from relativistic nuclear  collisions in recent years. Paquet ${\it {et \ al.}}$~\cite{kompost} have shown that the radiation from pre-hydro phase can be a substantial contribution to the photon and hadronic observables. They introduced a  pre-hydro phase KOMPOST between the CGC initial state and the hydrodynamical model evolution to study the effect  on photon spectra and elliptic flow parameter. The inclusion of this new phase enhances the photon production significantly which is reflected in  the $p_T>$  3 GeV region of the spectrum. However, 
the anisotropic flow parameter is found to change only marginally by the inclusion of the pre hydro phase.

The direct photon puzzle still remains as one of the interesting unsolved problems in relativistic heavy ion collisions. The experimental data  from a number of small and large systems as well as from different beam energies have opened up the possibility  of understanding the initial state as well as the  photon puzzle better.  Additionally, photon production from asymmetric collisions (C+Au)~\cite{phot_v1} and also from collisions of deformed nuclei (U+U)~\cite{uu} can play an important role in this regard. % in the determination of the photon anisotropic flow parameter. 

The fully overlapping U+U collisions can lead to different collision geometry depending on the orientation of the colliding nuclei.  A recent study using hydrodynamical model calculation has shown that the photon production from tip-tip configuration of U+U collisions is comparable to the production from most central  Au+Au collisions at RHIC~\cite{uu}. On the other hand, the elliptic flow from body-body configuration of uranium nuclei is found to be close to the photon $v_2$ from mid-central Au+Au collisions.  
The directed flow of photons from a hydrodynamical model calculation has  been found to be significantly large and it shows different behaviour compared to the elliptic and triangular flow parameters~\cite{phot_v1}. The $v_1(p_T)$ is found to be negative at smaller $p_T$ values unlike the higher order flow coefficients and it is also found to be dominated by the  QGP radiation in the entire $p_T$ region. % completely dominates the photon $v_1$ and it is found to be more sensitive to the initial parameter of the model calculation compared to the photon $v_2$ and $v_3$ parameters.

It has been shown in a recent study that the initial state nucleon shadowing in the Monte Carlo Glauber model increases the anisotropic flow of photons significantly. The effect of this initial state nucleon shadowing is found to be more prominent for photon observables compared to the hadronic observables~\cite{shadow}.

A radiative recombination model is used by Nonaka $\it {et \ al.}$~\cite{nonaka} to study photon production from heavy ion collisions. In a recombination model, the hadrons are formed by coalescence of valence quarks. The model is modified to allow photon emission and  processes such as $q \bar{q} \rightarrow \pi^0 \gamma$ are considered when the QGP hadronizes. They obtain an exponential $p_T$ distribution for the photons similar to a thermal distribution with an effective  temperature given by the (blue-shifted) recombination temperature.

The photon-jet correlations can play an important role in understanding the initial state produced in relativistic nuclear collisions. The photon jet transverse momentum imbalance and azimuthal correlations have been studied  using JETSCAPE framework in p+p and heavy ion collisions at LHC energy by Sirimanna ${\it et \ al.}$~\cite{sirimanna} . The JETSCAPE is a multistage framework which uses several modules to simulate different stages of jet propagation through the QGP medium. A significantly improved agreement with the photon data has been observed compared to the earlier calculations.

The photon HBT interferometry can be useful to get information about the spatio-temporal evolution of the system produced in relativistic nuclear collisions~\cite{dks_hbt}.
The feasibility of such study at LHC energy has been estimated by Garcia-Montero $\it {et \ al.}$~\cite{garcia_hbt} where they show that the HBT correlation at low $k_T$ ($<1$ GeV) can be statistically significant.

\begin{figure}
%\centerline{\includegraphics*[width=5.0 cm,clip=true]{4.eps}}
\centerline{\includegraphics*[width=9.0 cm,clip=true]{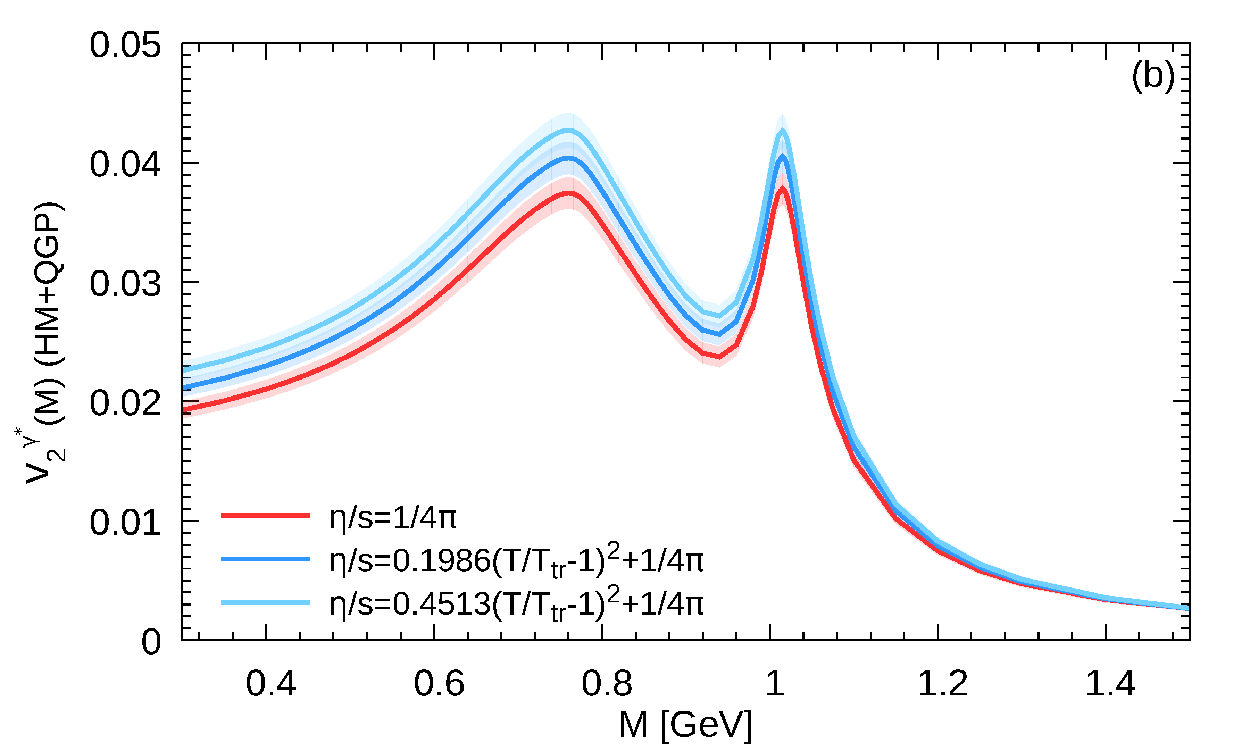}}
\caption{(Color online) Elliptic flow of thermal dileptons is sensitive to the shear viscosity parameter $\eta/s(T)$. The figure is from Ref.~\cite{gojko}.}
\label{fig2}
\end{figure}

\section{Dileptons}
The virtual photons or dileptons are also produced from all stages  of the fireball evolution similar to the real photons and are considered as a potential probe to study the properties of the strongly interacting matter produced in  relativistic nuclear collisions~\cite{wong, rapp}. The dileptons are massive unlike the the real photons. The invariant mass $M_{\rm {ll}}$ of the dilepton pair and the transverse momentum $p_T$, both are tuned to access the different stages of the matter evolution.

%The  dilepton pairs having large $M_{\rm {ll}}$ and $p_T$ are emitted from the hot and dense early stages of the collisions and those having relatively smaller $p_T$ and $M_{ll}$ are emitted from the later stages of the evolving system when the flow is strong.
%A schematic of the dilepton mass spectrum is shown in Fig.~\ref{dil}. 

In a dilepton mass spectrum, the region below $\phi$ mass is known as the low mass region, $\phi$ mass to $J/\Psi$ mass is known as intermediate mass region and above that is called the high mass region. The  dilepton pairs having large $M_{\rm {ll}}$ and $p_T$ are mostly emitted from the hot and dense early stages of the collisions and those having relatively smaller $p_T$ and $M_{\rm ll}$ are emitted from the later stages of the evolving system when the flow is strong.

It has been shown that dileptons with $M_{\rm ll} >$ 1 GeV are dominated by QGP radiation whereas, $M_{\rm ll} <$ 1 GeV dileptons are mostly due to hadronic matter radiation and are important to study the chiral symmetry breaking/restoration. The anisotropic flow of dileptons using a hydrodynamical model calculation shows rich structure  as a function of  $p_T$ as well as invariant mass due to the interplay of the emission from fluid elements at different temperatures with varying radial  flow pattern~\cite{rupa_dil_v2}. 

Recent studies have shown that Bayesian analysis simulating the soft hadronic observables from various stages of heavy ion collisions can be useful to constrain the transport coefficients. The preliminary results by JETSCAPE simulation group show that the high temperature behaviour of the  $\eta/s$ and $\zeta/s$ are not very clear and the anisotropic flow of dileptons can be used to constrain these  shear and bulk viscosity coefficients~\cite{gojkohp}.
Transport models such as coarse grained URQMD~\cite{urqmd} and SMASH~\cite{smash} are used to study the dilepton production at lower beam energies. Vujanovic ${\it et \ al.}$~\cite{gojko} have shown  that the dilepton $v_2$ at large invariant masses can be a probe for QGP and the results are sensitive to the relaxation time.

The new results of dilepton production from photon-photon interactions  have gathered a lot of attention in recent times~\cite{gg}.
The intense electromagnetic field surrounding a highly charged   heavy nucleus in relativistic collisions provides a flux of quasi-real photons where the photon flux increases with (square of) nuclear charge. The dileptons  produced through the $\gamma \gamma \rightarrow l^+ l^-$ process are measured in ultra-peripheral collisions. %The photon flux increases with (square of) nuclear charge  (z$^2$) and the photoproduction is distinctly peaked at low $p_T$.
The STAR and ATLAS experiments  have also measured the dilepton production from photon interactions in hadronic collisions complementing the results from  ultra-peripheral collisions.

The new low mass preliminary STAR dielectron data  from Au+Au collisions at 27 and 54.4 GeV  significantly enhance the precision of the in medium rho modification measurements compared to the STAR BES-I results~\cite{star_ee}.  The ALICE Collaboration has reported  high statistical precision dielectron spectra from Pb+Pb and p+p collisions at 5.02A TeV at LHC  analyzing the 2018 data and also soft dielectrons from p+p collisions at 13 TeV with B=0.2 T~\cite{alice_ee}. 
%The ALICE dielectron data from 5.02A TeV  Pb+Pb collisions and for different centrality bins obtained using 2018 Pb+Pb data provide higher statistical  precision compared to previous measurements.

The STAR experiment at RHIC has reported the invariant mass and yield distribution of inclusive dimuons in the low $p_T$ region  for different centralities of Au+Au collisions. The measurement is done in the mass range 3.2 to 10 GeV/$c^2$ utilizing the Muon Telescope Detector. They observe a significant enhancement with respect to the cocktail in the 60--80\% centrality bin and data are found to be consistent with theoretical calculations~\cite{star_mu}.

Interesting new data on the impact parameter dependence of dimuon acoplanarity in ultra-peripheral Pb+Pb collisions have been reported by the CMS Collaboration. The acoplanarity is basically the relative angular deflection of the dimuon pair.  They show that the centrality dependent $\gamma \gamma \rightarrow \mu^+ \mu^-$  production provide valuable insight about the origin of observed broadening of lepton pairs produced from $\gamma \gamma$ scatterings in hadronic collisions~\cite{cms_mu}. % while rapidity dependence constrains the relative contributions from leading order and high order photon photon interaction to measure dimuon pairs.

The yield and distributions of dimuons from $\gamma \gamma \rightarrow \mu^+ \mu^-$ processes in Pb+Pb collisions have been measured by ATLAS collaboration in ultra-peripheral as well as non-ultra peripheral collisions. The distribution of acoplanarity and $k_T$ show significant centrality dependence in the preliminary ATLAS  data~\cite{atlas_mu}.

%The non-UPC production of dimuons from two photon scattering in Pb+Pb collisions with the ATLAS detector show that muon pairs produced via two photon scattering processes  in hadronic Pb+Pb collisions provide a potentially sensitive  electromagnetic probe of the QGP. 

%A new experiment NA60+ at CERN SPS has been proposed which  will access the high $\mu_B$ region of the QCD phase diagram using fixed target experiment~\cite{na60}. 
\section{W{$^\pm$} and $Z$ bosons}

\begin{figure}
\centerline{\includegraphics*[width=9.0 cm,clip=true]{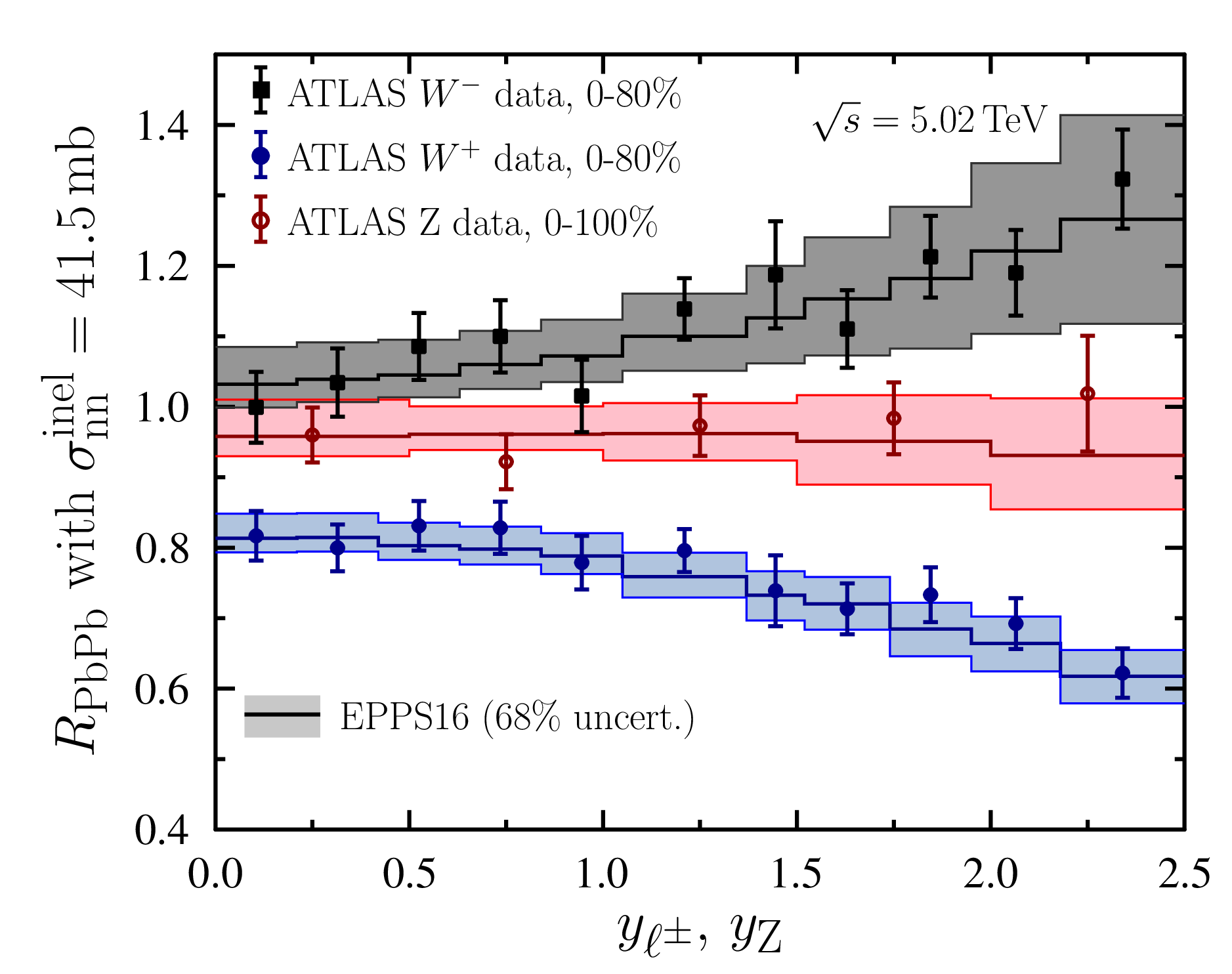}}
\caption{(Color online) The nuclear modification factor $R_{\rm PbPb}$ as a function of rapidity calculated using nuclear-suppressed $\sigma^{\rm inel}_{\rm nn}=$41.5 mb from Ref.~\cite{kari}. ATLAS data points are from~\cite{at1, at2}.}
\label{fig3}
\end{figure}

The massive vector bosons are produced from the initial state in heavy ion collisions before the QGP is formed and are considered as valuable probes to study the nuclear modifications of the parton distribution functions (PDF). 

The  effect of nuclear shadowing on the inelastic nucleon nucleon cross-section $\sigma_{\rm {NN}}$ has been studied by Eskola $\it {et. al.}$~\cite{kari} in an interesting work using the high precision ATLAS $W^\pm$ and $Z$ data from  Pb+Pb collisions at 5.02A TeV~\cite{at1, at2}. The Glauber model formalism is widely used in centrality dependent  calculations in heavy ion collisions where the beam energy dependence is incorporated using  the $\sigma_{\rm {NN}}$  obtained from the p+p measurements. Thus, a suppression in the value of $\sigma_{\rm {NN}}$ is expected  due to shadowing and saturation phenomena at small $x$ values for larger nuclei compared to protons. The NNLO QCD calculations and nuclear PDFs are used to estimate the $\sigma_{\rm NN}$ in~\cite{kari} and a significantly suppressed value (about 41.5 mb down from 71 mb at 5.02A TeV)  is obtained which is expected to affect the experimental analysis of the nuclear modification factor.

Gauge boson associated with jet production can provide valuable information about jet-quenching and parton energy loss in the hot and dense medium. The  $W^\pm$+jets production at LHC energy has been studied by Zhang ${\it et \ al.}$~\cite{sherpa} using a Monte Carlo event generator SHERPA (a sharp and smooth algorithm), the Linear Boltzmann transport model and a parton shower. The $W^\pm$+jets, which are dominated by quark jets are expected to provide complementary information on jet quenching along with Z+jets and $\gamma$-jets.
 
There has been significant development in the $W^\pm$ and $Z$ boson measurements from p+p, p+Pb, and Pb+Pb collisions at different LHC energies. The ATLAS Collaboration has measured the $W^\pm$ and Z bosons  from p+p and Pb+Pb collisions at 5.02A TeV  where the production yields are observed via their leptonic decay channels~\cite{atlas_w}. The p+p data are found to be in good agreement with the NNPDF3.1 and the Pb+Pb data match well with all the theory predictions. They also show that the isospin  effect plays an important role in the analysis of $W^\pm$ data. 

The CMS Collaboration has recently reported preliminary $Z$ boson data from Drell-Yan process at 8.16A TeV p+Pb collisions at LHC and the new measurement is extended to a lower mass region where the data are found to be explained well by EPPS16 with shadowing compared to free nucleon PDFs~\cite{cms_w}. The Z boson yield at various centrality bins are compared to the HG-PYTHIA model. Additionally, a high precision $Z$ boson azimuthal anisotropy measurement from Pb+Pb collisions by CMS Collaboration shows that the $v_2$ is consistent with zero.

%The Drell-Yan measurement by CMS is extended to a lower mass region and the data is found to be explained well by EPPS16 with shadowing compared to free nucleon PDF.
The production of $Z$ boson from 8.16A TeV p+Pb collision is also reported by LHCb experiment where the predictions at the forward and backward rapidities are found to be sensitive to the nPDFs in a unique kinematic domain~\cite{lhcb}. The structure of the nucleus can be studied in a complementary fashion from the LHCb results.

The ALICE  W$^\pm$   data  from Pb+Pb collisions at 5.02A TeV show that the $R_{\rm {AA}}$  does not depend on the collision centrality significantly.  The $Z$ boson $R_{\rm {AA}}$ by ALICE is found to be consistent with theoretical calculations and only at large rapidities the value deviates from 1~\cite{alice_w}.  %The ALICE data show that the $R_{\rm {AA}}$ for  W$^\pm$ does not show any centrality dependence.

All these new high precision data and the sophisticated theory calculations promise a bright future for the electroweak probes in relativistic nuclear collisions.


\begin{thebibliography}{99}

\bibitem{mclaurren} L. D. McLerran and T. Toimela, 
%Photon and Dilepton Emission from the Quark - Gluon Plasma:  Some General Considerations,
\ Phys. \ Rev. \ D ${\bf 31}$, 545 (1985).

\bibitem{gojkohp} G. Vujanovic, arXiv:2008.11843.

\bibitem{zvi} Z. Citron, in these proceedings (2020).

\bibitem{dks_qm}D. K. Srivastava, \ J. \ Phys. \ G ${\bf 35}$, 104026 (2008); R.  Chatterjee, L.  Bhattacharya, D. K. Srivastava, \ Lect. \ Notes \ Phys. ${\bf 785}$  219 (2010).
\bibitem{wa98} M. M. Aggarwal ${\it et \ al.}$ [WA98 Collaboration],  \ Phys. \  Rev. \  Lett. ${\bf 85}$, 3595 (2000).


\bibitem{ph_g}] A. Adare ${\it et \ al.}$ [PHENIX Collaboration] \ Phys. \ Rev. \ Lett. ${\bf 104}$, 132301 (2010); A. Adare ${\it et \ al.}$ [PHENIX Collaboration] \ Phys. \ Rev. \ C ${\bf 91}$, 064904 (2015).

\bibitem{al_g} M. Wilde for the ALICE Collaboration, \ Nucl. \ Phys. ${\bf A904}$ , 573c (2013); J. Adam ${\it et \ al.}$ [ALICE Collaboration], \ Phys. \ Lett. \ B ${\bf 754}$, 235 (2016).

\bibitem{v2_ex} A. Adare ${\it et \ al.}$ [PHENIX Collaboration]. \ Phys. \ Rev. \ Lett. ${\bf 109}$, 122302 (2012); A. Adare ${\it et \ al.}$ [PHENIX Collaboration], \ Phys. \ Rev.  \ C ${\bf 94}$, 064901 (2016).

\bibitem{v2_ex1} D. Lohner for the [ALICE Collaboration], \ J. \ Phys. \ Conf. \ Ser. ${\bf 446}$, 012028 (2013).
\bibitem{phot_v2}    R. Chatterjee, E. S. Frodermann, U. W. Heinz, D. K. Srivastava,  \ Phys. \ Rev. \ Lett. ${\bf  96}$  202302 (2006).
\bibitem{puzzle} R. Chatterjee, H. Holopainen, I. Helenius, T. Renk, K.J. Eskola, \ Phys. \ Rev. \ C ${\bf 88}$, 034901 (2013).
\bibitem{veronika} V. C. Roman, in these proceedings (2020).

\bibitem{phot_scaling} A. Adare ${\it et \ al.}$, \ Phys. \ Rev. \ Lett. ${\bf 123}$, 022301 (2019).

\bibitem{dks} J. Cleymans, K. Redlich, and D. K. Srivastava,  \ Phys. \ Lett. \ B ${\bf 261}$, (1998).
\bibitem{Dhrub} D. Dixit, in these proceedings (2020).
\bibitem{kompost} J.-F. Paquet, C. Shen, B. Schenke, and C. Gale, in these proceedings (2020) ;. Gale, J.-F. Paquet, B. Schenke, and C. Shen,  arXiv:2002.05191.

\bibitem{phot_v1}  P. Dasgupta, R. Chatterjee, and D. K. Srivastava,  J. Phys. G ${\bf 47}$  085101 (2020).


\bibitem{uu} P. Dasgupta, R. Chatterjee, and D. K. Srivastava, Phys. \ ReV. \ C ${\bf 95}$, 064907 (2017).
\bibitem{shadow} P. Dasgupta, R. Chatterjee, S. Singh, and J. Alam, \ Phys. \ Rev. \ C ${\bf 97}$, 034902 (2018).
\bibitem{nonaka} H. Fujii, K. Itakura, C. Nonaka, \ Nucl. \ Phys. ${\bf A 967}$, 704 (2017).

\bibitem{sirimanna} C. Sirimanna ${\it et \ al.}$ [JETSCAPE Collaboration], in these proceedings (2020).

\bibitem{dks_hbt} D. K. Srivastava, \ Phys. \ Rev. \ C ${\bf 71}$, 034905 (2005).
\bibitem{garcia_hbt} O. Garcia-Montero ${\it et \ al.}$ \ Phys. \ Rev. \ C ${\bf 102}$, 024915 (2020).


\bibitem{wong} C. Y. Wong,{\it Introduction of High Energy Heavy Ion Collisions}, World Scientific, Singapore, 1994.
\bibitem{rapp} R. Rapp, \ Phys. \ Rev. \ C ${\bf 63}$, 054907 (2001).
\bibitem{rupa_dil_v2} R. Chatterjee, D. K. Srivastava, U. Heinz and C. Gale,  \ Phys. \ Rev. \ C ${\bf 75}$, 054909 (2007).

\bibitem{urqmd} S. Endres, H. van Hees, and M. Bleicher, \ Phys. \ Rev. \ C ${\bf 94}$, 024912 (2016).
\bibitem{smash} J. Staudenmaier ${\it et \ al.}$, \ Phys. \ Rev. \ C ${\bf 98}$, 054908 (2018).
\bibitem{gojko} G. Vujanovic, G. S. Denicol, M. Luzum, S. Jeon, and C Gale,  \ Phys. \ Rev. \ C ${\bf 98}$, 014902 (2018).

\bibitem{gg}S. Klein ${\it et \ al.}$ \ Phys. \ Rev.  \ Lett. ${\bf 122}$, 132301 (2019).
\bibitem{star_ee} Z. Wang [STAR Collaboration], in these proceedings (2020).
\bibitem{alice_ee} D. Sekihata, in these proceedings (2020); S. Acharya ${\it et \ al.}$, [ALICE Collaboration] arXiv:2005.14522.
\bibitem{star_mu} Z. Ye, [STAR Collaboration], in these proceedings (2020).
\bibitem{cms_mu} S. Yang [CMS Collaboration], in these proceedings (2020).
\bibitem{atlas_mu} S. Mohapatra [ATLAS Collaboration], in these proceedings (2020).


\bibitem{kari} K. J. Eskola, I. Helenius, M. Kuha, and H. Paukkunen, arXiv:2008.13448; K J. Eskola, I. Helenius, M. Kuha, H. Paukkunen, arXiv:2003.11856.
\bibitem{at1} G. Aad ${\it et al.}$ [ATLAS collaboration],\ Eur. \ Phys. \ J. \ C ${\bf 79}$  935 (2019). 
\bibitem{at2} G. Aad ${\it et al.}$ [ATLAS collaboration], \ Phys. \ Lett. \ B ${\bf 802}$, 135262(2020).


\bibitem{sherpa} S.-L. Zhang, in these proceedings, (2020).


%\bibitem{na60} E. Scomparin, in these proceedings (2020); M. Agnello {\it et al.}, arXiv:1812.07948.
 



\bibitem{atlas_w}  I. Grabowska-Bold, in these proceedings (2020).
\bibitem{cms_w} A. A. Baty, in these proceedings (2020).
\bibitem{lhcb} Hengne Li, in these proceedings (2020).
\bibitem{alice_w} G. Taillepied,  in these proceedings (2020).
%\bibitem{fcc} P. Dasgupta, S. De, R. Chatterjee, and D. K. Srivastava, \ Phys. \ Rev. \ C {\bf 98}, 024911 (2018).







\end{thebibliography}
\end{document}